# Contextualizing the global relevance of local land change observations

**N. R. Magliocca (1), E. C. Ellis (1), T. Oates (2) and M. Schmill (2)**

(1) Department of Geography and Environmental Systems, University of Maryland, Baltimore County, Baltimore, Maryland, USA
(2) Department of Computer Science and Electrical Engineering, University of Maryland, Baltimore County, Baltimore, Maryland, USA

Email: nmag1@umbc.edu

**Abstract**
To understand global changes in the Earth system, scientists must generalize globally from observations made locally and regionally. In land change science (LCS), local field-based observations are costly and time consuming, and generally obtained by researchers working at disparate local and regional case-study sites chosen for different reasons. As a result, global synthesis efforts in LCS tend to be based on non-statistical inferences subject to geographic biases stemming from data limitations and fragmentation. Thus, a fundamental challenge is the production of generalized knowledge that links evidence of the causes and consequences of local land change to global patterns and vice versa. The GLOBE system was designed to meet this challenge. GLOBE aims to transform global change science by enabling new scientific workflows based on statistically robust, globally relevant integration of local and regional observations using an online social-computational and geovisualization system. Consistent with the goals of Digital Earth, GLOBE has the capability to assess the global relevance of local case-study findings within the context of over 50 global biophysical, land-use, climate, and socio-economic datasets. We demonstrate the implementation of one such assessment - a representativeness analysis - with a recently published meta-study of changes in swidden agriculture in tropical forests. The analysis provides a standardized indicator to judge the global representativeness of the trends reported in the meta-study, and a geovisualization is presented that highlights areas for which sampling efforts can be reduced and those in need of further study. GLOBE will enable researchers and institutions to rapidly share, compare, and synthesize local and regional studies within the global context, as well as contributing to the larger goal of creating a Digital Earth.

## 1. Introduction
Much of our knowledge of global environmental change, and in particular its connection to land use, has been synthesized from local observations. These local observations, in the form of case studies, are considered to be the gold standard for investigating the causes and consequences of local land-use change [1]. For this reason, land change scientists must generalize globally from case-study observations made locally and regionally to understand land use as a global change process. Meta-studies across sets of local case studies have emerged as one of the primary methods used in land change science to research regional to global patterns in the causes and consequences of local land change [2]. However, the case-studies upon which meta-studies rely are not conducted at random locations across Earth's land. As a result, a sample of existing case studies can be highly biased, over-representing or under-representing more accessible areas, wealthy areas, high population areas, the temperate zone, etc., which can compromise the broader-scale inferences of the meta-study [3].

The GLOBE project was developed in part to address this issue. The GLOBE project aims to transform land change science by enabling new scientific workflows built on statistically robust global integration of local and regional observations using a social-computational system available freely online. The computational system of GLOBE can assess the degree to which a given collection of case studies is an unbiased sample across an extent of Earth's land based on one or several global variables of interest, and to potentially remediate bias in case-study samples by identifying underrepresented areas. This is done through *representativeness* and *representedness* analyses designed primarily to assist researchers conducting meta-studies of existing local or regional case studies by detecting under- and over-studied areas.

We first discuss the related concepts of representativeness and representedness, and describe the algorithm implemented in the GLOBE system to carry out the analyses. Both analyses are then demonstrated with a recently published meta-study of global changes in swidden cultivation by van Vliet and colleagues [4]. A set of experiments assess how well the collection of case studies used in the meta-study represents global patterns in percent tree cover and population density, as well as comparing analyses over two different spatial extents that more or less precisely encompass the global area of interest. Findings from these analyses are presented in a series of screen shots from the GLOBE online system. Limitations of the current system are discussed, followed by a description of future capabilities designed to further advance synthesis in land change science.

## 2. Methods
### 2.1. Representativeness and representedness
*Representativeness* analysis assesses the degree to which a given collection (i.e. sample) of study sites represents an unbiased sample of a specified global extent (i.e. population) with respect to a set of global variables selected by the user. Representativeness can be assessed in two ways: distributional or variational representativeness. Distributional representativeness is a measure of how well the frequency distribution of the sample (i.e. collection of case-study sites) aligns with that of the population (i.e. specified global extent) with respect to a particular variable. Variational representativeness is an assessment of how well the variability within the sample covers the range of variability observed across the population. Since the objective of most meta-studies in land change science is to capture the central tendencies of land change trends [2], we will consider distributional representativeness for these analyses.

Representativeness analysis is based on the principle that unbiased samples of study sites would characterize the distribution of variation observed in a global variable(s) to the same degree as a random sample of the same size. The degree to which a sample collection is representative of a specified global extent is quantified by how similar its frequency distribution(s) are to that of the entire specified global extent (one of several *f*-divergence indicators can be used, [5]). The representativeness of the sample (measured as a single indicator value from 0 to 1) is then compared to the frequency distribution of representativeness indicator values for a large set of automatically generated random samples. From this, the probability of attaining a given level of representativeness by random sampling can be compared against the representativeness of the collection; samples with representativeness levels similar to the central 50% of random samples are not significantly biased, and those with lower levels show bias.

*Representedness* is a related concept that assess how well locations within a specified global extent are represented by the collection of study sites with respect to a set of global variables. Again, the frequency distributions of the sample and population for a given variable are compared. If the frequency of data values of the sample - obtained from the specified global dataset at each case-study site location - is greater than that of the population at any value, then case-study sites with that value are over-represented in the sample, and vice versa. Differences in frequency distributions at each variable value are normalized between values of -1 and 1 for perfectly under- and over-represented, respectively, with a value of 0 implying a well-represented variable value. The representedness of areas within the specified global extent are color-coded as a heat map for quick geovisualization, and the global extents of well-represented, under-represented and over-represented areas are calculated as a total area (km$^2$) and as a percentage of the specified global extent.

*2.2. Data Structure*
The global analytic capabilities of GLOBE are achieved by stratifying Earth's land surface into a set of equal-area hexagon tiles derived using the geodesic discrete global grid (DGG) system of Sahr [4], [5]. The full set of roughly 1.44 million 96 km$^2$ hexagonal ISEAA3H Level 12 DGG tiles covering Earth's land surface serve as the foundational units for global analysis in GLOBE (GLOBE Land Units; GLU). In the GLOBE system, global variable values are recorded for each GLU, thereby enabling rapid calculation of area-weighted statistics across the entirety of Earth's land surface.

A growing set of publically available global datasets are integrated into the GLOBE system. The full set and descriptions of each variable can be found at http://globe.umbc.edu/documentation/global-variables/. Various geoprocessing methods are used in ArcGIS 10.0 (ESRI, Redlands, CA) to convert global variable datasets from their native formats to the GLUs used in GLOBE. First, each global variable is converted to 30 arcsec raster with an extent equal to LandScan 2007 [8], which is used as the base layer. Global variables that have smaller extents than LandScan 2007 are first extrapolated in their native resolutions using the 'Focal Statistics' function before being converted to 30 arcsec resolution. Once converted to 30 arcsec resolution, all datasets are processed with the DGG raster (30 arcsec resolution) using the 'Zonal Statistics' tool to obtain values for each GLU.

*2.3. Experiments*
Three experiments are conducted with a recently published meta-study of changes in swidden agriculture from van Vliet et al. [3] to test the representativeness and representedness algorithms. The first analysis is performed with the van Vliet collection based on global percent tree cover from MODIS 2003 data [9]. Since the van Vliet meta-study investigates swidden agriculture, which is typically found in tropical forests, we expect that the collection will not be representative of percent tree cover over the entire global extent. We then repeat this analysis with a more precise spatial constraint by filtering the global extent to include just the tropical woodland biomes based on global potential vegetation classes [10]. Limiting the extent of the analysis to tropical woodlands is more consistent with the intended study area of the van Vliet et al. [4] meta-study, which should be reflected by a bias indicator value closer to that of a random sample. However, because swidden agriculture often entails deforestation, we expect that the van Vliet collection is still biased towards less tree cover than that observed across all tropical forests. Thus, we repeat the representativeness analysis, limited to tropical forests, but based on global population density, to assess whether the collection is more representative of population densities observed across tropical forest locations.

Screen shots of the representativeness analyses performed in GLOBE are provided below. The input parameters of the analysis are seen on the left side of the screens. Effective sample size is set to equal the number of cases in the van Vliet collection (n=157), which also specifies the size of each of the 1000 random samples taken from the within the area of analysis (i.e. population). Histograms of the van Vliet collection and population are shown at the bottom of the screens and illustrate the percentage of GLUs with the specified global variable value (fig. 1). The third histogram displays the bias indicator value of the collection (red circle) compared to the distribution of bias indicator values generated from the random samples of the population (fig. 1). The map is color coded by representedness values, with dark red as 'very under-represented', green as 'well represented', and dark blue as 'very over-represented'.

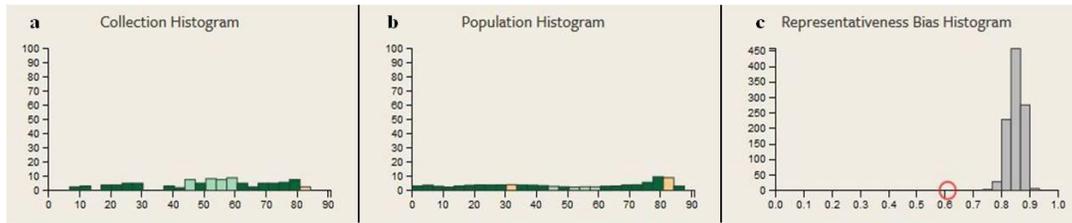

Figure 1: Histograms of the (a) collection, (b) population of GLUs in the analysis extent, and (c) bias indicator values from random samples (gray bars) and collection (red circle).

## 3. Results

The first representativeness analysis of the van Vliet collection was conducted over the entire global extent and based on percent tree cover (fig. 2). The collection was highly biased (representativeness bias = 0.4) in comparison to a random sample (mean = 0.75), as low percent tree cover areas (e.g. desert and arctic regions) were under-represented (red) and some forested areas were slightly over-represented (light green). Figure 3 shows a revised representativeness analysis that was limited to tropical forests, taking into account the tendency for swidden agriculture to be limited to tropical areas. As a result, the collection appeared less biased (0.62) in comparison to a random sample (mean = 0.85). Areas of moderate tree cover were slightly over-represented, while areas with high percentages of tree cover (e.g. central Amazon) were under-represented. The final representativeness analysis used the same spatial extent as the previous analysis, but was instead based on population density (fig. 4). With respect to population density, the van Vliet collection was roughly as unbiased (0.75) as a random sample (mean = 0.8). Most tropical forest areas were well represented by the collection, with the exception of parts of the Amazonian region which were slightly under-represented.

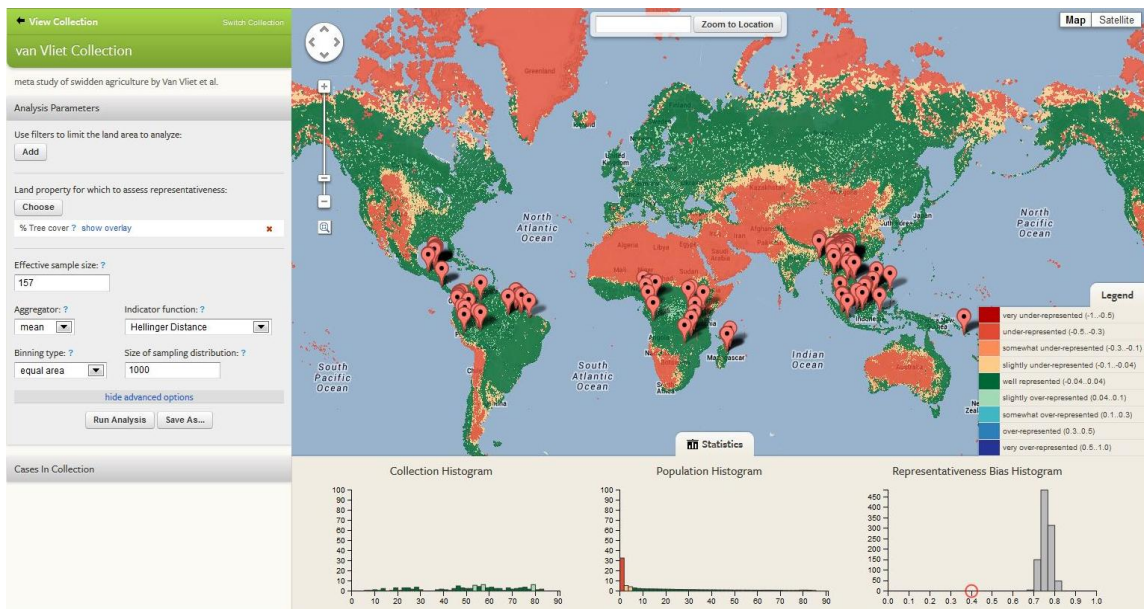

Figure 2: A screen shot from a GLOBE representativeness analysis of the collection of sites contained in van Vliet et al. [4] based on global percent tree cover. Dark red is 'very under-represented', green is 'well represented', and dark blue is 'very over-represented'.

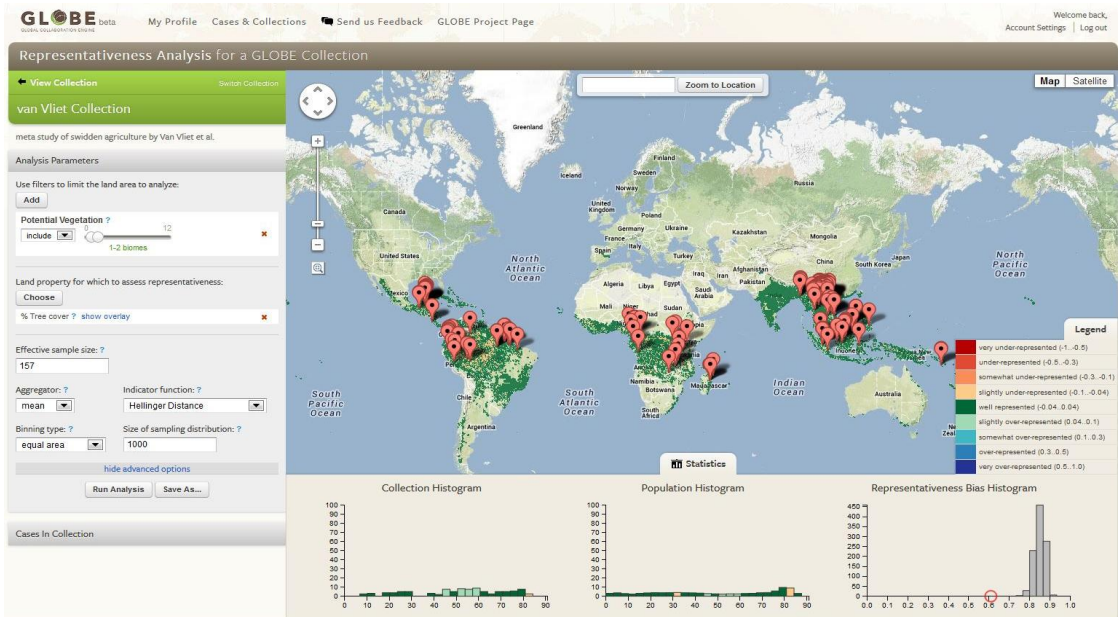

Figure 3: Representativeness analysis of the van Vliet collection based on global percent tree cover limited to the spatial extent of tropical biomes.

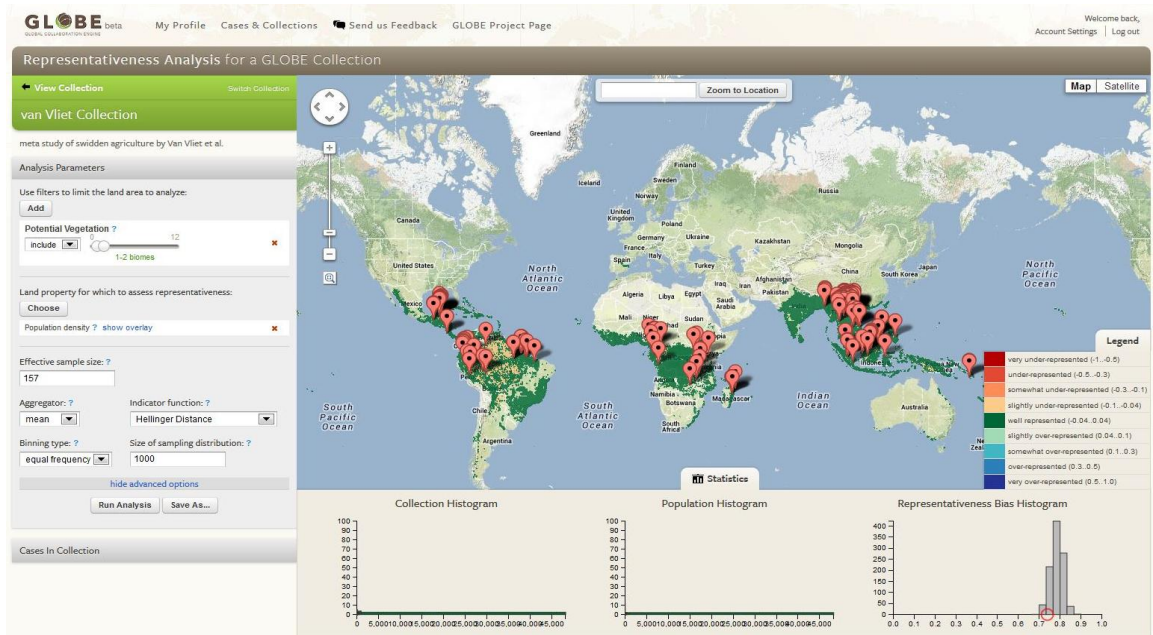

Figure 4: Representativeness analysis of the van Vliet collection based on global population density limited to the spatial extent of tropical biomes.

## 4. Discussion and conclusions

The results of the three representativeness analyses illustrated the importance of quantifying and bounding the areas of the Earth's surface to which meta-study findings apply. The van Vliet collection was found to be a slightly biased representation of tropical forest areas based on percent tree cover. However, this was to be expected given that swidden agriculture takes place in close proximity to populated areas and implies a degree of deforestation. In fact, the collection was found to be a nearly unbiased sample of tropical forest sites based on population density, and thus the meta-study findings are statistically representative of other tropical forest sites with similar population densities beyond the collection of cases analyzed.

The GLOBE system advances current global assessment and synthesis efforts by providing a spatially explicit, quantitative assessment of the representativeness of georeferenced local observations in the context of global biophysical and socio-economic variables. This enables statistically robust global inferences to be made from local observations, which is necessary to characterize the generalizability of meta-study findings. The representativeness analyses in GLOBE are still being developed, and are currently limited to a single variable. Future versions will include functionality for multivariate representativeness analysis. Additional future capabilities will provide users with tools to improve the representativeness of existing collections. This can be done by either calculating a set of weights to compensate for over- and under-represented case studies, and/or searching for case studies specifically within underrepresented areas to add to a collection. More broadly, GLOBE will enable researchers and institutions to rapidly share, compare, and synthesize local and regional studies within the global context, as well as contributing to the larger goal of creating a Digital Earth.

**Acknowledgements**
The GLOBE project is supported by the US National Science Foundation under grant NSF # 1125210. Any opinions, findings, and conclusions or recommendations expressed in this material are those of the authors and do not necessarily reflect the views of the National Science Foundation.